\documentclass[a4paper,11pt]{article}
 \usepackage[margin=1in]{geometry}
\usepackage[square,numbers]{natbib}
\usepackage[utf8]{inputenc}
\usepackage{amsmath}
\usepackage{amsthm}
\usepackage{amsfonts}
\usepackage{bm}
\usepackage{thm-restate}
\usepackage{wrapfig}
\usepackage{subcaption}
\usepackage{float}
\usepackage{algorithm}
\usepackage[noend]{algorithmic}

\usepackage{color}   
\usepackage{hyperref}
\hypersetup{
    linktoc=all,     
    linkcolor=blue,  
}

\usepackage[group-separator={,}]{siunitx}
\usepackage{framed}
\usepackage{setspace}

\usepackage{todonotes}

\theoremstyle{plain}

\theoremstyle{definition}

\theoremstyle{remark}

\usepackage{dsfont}

\newcommand{\remove}[1]{}

\allowdisplaybreaks

\begin{document}
\pagenumbering{arabic}

\title{
An Efficient Simulation-Based Travel Demand Calibration Algorithm for Large-Scale Metropolitan Traffic Models\footnote{authors ordered alphabetically.}
}
\date{}

 \author{
 Neha Arora \\
 Google Research \\
 \small\texttt{nehaarora@google\!.\!com}
 \and
Yi-fan Chen\\
 Google Research \\
 \small\texttt{yifanchen@gmail\!.\!com}
 \and
Sanjay Ganapathy\\
 Google Research \\
 \small\texttt{ganapathys@google\!.\!com}
 \and
Yechen Li\\
 Google Research \\
 \small\texttt{yechenl@google\!.\!com}
 \and
Ziheng Lin\\
 Google Research \\
 \small\texttt{zihenglin@google\!.\!com}
 \and 
Carolina Osorio\\
 Google Research \\
 HEC Montreal \\
 \small\texttt{osorioc@google\!.\!com}
 \and
Andrew Tomkins\\
 Google Research \\
 \small\texttt{tomkins@google\!.\!com}
 \and
Iveel Tsogsuren\\
 Google Research \\
 \small\texttt{iveel@google\!.\!com}
 \and
}

\maketitle

\begin{abstract}
Metropolitan scale vehicular traffic modeling is used by a variety of private and public sector urban mobility stakeholders to inform the design and operations of road networks. High-resolution stochastic traffic simulators are increasingly used to describe detailed demand-supply interactions. The design of efficient calibration techniques remains a major challenge. This paper considers a class of high-dimensional calibration problems known as origin-destination (OD) calibration. We formulate the problem as a continuous simulation-based optimization problem. Our proposed algorithm builds upon recent metamodel methods that tackle the simulation-based problem by solving a sequence of approximate analytical optimization problems, which rely on the use of analytical network models. In this paper, we formulate a network model defined as a system of linear equations, the dimension of which scales linearly with the number of roads with field data  and independently of the dimension of the route choice set. This makes the approach suitable for large-scale metropolitan networks. The approach has enhanced efficiency compared with past metamodel formulations that are based on systems of nonlinear, rather than linear, equations. It also has enhanced efficiency compared to traditional calibration methods that resort to simulation-based estimates of traffic assignment matrices, while the proposed approach uses analytical approximations of these matrices. We benchmark the approach considering a peak period Salt Lake City case study and calibrate based on field vehicular count data. The new formulation yields solutions with good performance and is suitable for large-scale road networks. 
\end{abstract}
{\bf Keywords:} traffic simulator calibration, simulation-based optimization, travel demand estimation

\section{Introduction}
\label{sec:intro}
Cities, as well as various private sector urban mobility stakeholders,  increasingly use detailed urban mobility models to inform the design of mobility services and policies (e.g., congestion pricing, ride-sharing, car-sharing). The most detailed of these models are stochastic traffic simulation models known as microscopic  simulators. They provide a disaggregate description of demand by representing individual vehicles, travelers and trips. They also provide a detailed representation of supply by, for instance, embedding a description of traffic control strategies, such as traffic lights and variable speed signs.  However, this increased modeling resolution of both demand and supply comes with an increase in the number of  input parameters that need to be estimated or calibrated, as well as an increase in the compute time needed to evaluate the model. Hence, there is a need for calibration algorithms that perform well within a small data context, i.e., algorithms that  identify well performing solutions within few (simulation) function evaluations. 
Arguably, the most difficult and important offline calibration problem is that of estimating travel demand. This calibration problem is known as origin-destination (OD) calibration. Traditionally, the region of interest (e.g., metropolitan area, city center) is segmented into a set of discrete zones, known as Travel Analysis Zones (TAZes). A set of feasible pairs of TAZes,  known as  origin-destination pairs, is assumed to be known. The static OD calibration problem then consists of estimating, for a given time interval, the expected travel demand between these OD pairs. It's dynamic counterpart estimates the expected travel demand over a set of discrete time-intervals. In this paper, we focus on the static OD calibration problem.   A recent review of the OD calibration literature is in \cite{Osorio19a}.

This problem is a continuous high-dimensional simulation-based optimization (SO) problem. The dimension of the problem is equal to the number of OD pairs. In metropolitan areas, this number can be in the order of thousands to tens of thousands. In this paper, we consider a problem of dimension in the order of thirty thousand. The problem is typically tackled by using general-purpose SO algorithms and terminating them when a given computational budget is depleted, i.e., when an upper bound on the number of function evaluations  or on the total compute time is reached. In the literature, the computational budgets used are 1 to 2 orders of magnitude lower than the problem dimension.  

The most popular algorithms used to tackle OD calibration problems are Simultaneous Perturbation Stochastic Approximation (SPSA) \cite{Spall92,Spall03} such as in \cite{Balakrishna15, Vaze09, Lee09, BenAkiva12}, and genetic algorithms  such as in \cite{Kim01, Stathopoulos04, Vaze09}. However, these general-purpose algorithms are designed based on asymptotic performance guarantees and are not designed to be terminated with such few function evaluations. Specific shortcomings of these commonly used methods when used for OD calibration are well documented \cite{Antoniou15, Tympakianaki15, Tympakianaki18}, and have led to recent extensions of SPSA, such as in \cite{Cipriani11, Lu15, Antoniou15, Tympakianaki15}.

The lack of SO algorithms that can  tackle these high-dimensional OD calibration problems efficiently is acknowledged in the literature \cite{Djukic14_pg33}, and more so in practice. Advances in this area are also important building blocks to tackle real-time calibration problems \cite{Bierlaire04, Zhou07, Barcelo15}.
This paper tackles this challenge. It focuses on the design of SO algorithms that preserve asymptotic convergence properties while also performing well when few function evaluations are allowed.

A recent approach to reduce function evaluations required for optimization is to use an analytic metamodel that can approximate the simulation response that is computationally far more efficient. Metamodel techniques have been successfully applied to various classes of continuous and discrete transportation SO problems, including signal control, car-sharing fleet assignment, and toll optimization  \cite{Osorio13,Osorio15a,Zhou17,Osorio21}. They have been recently formulated for low-dimensional car-following calibration problems \cite{Osorio19c} and for high-dimensional OD calibration problems \cite{Osorio19a,Osorio19b}. This paper builds upon the metamodel SO work for OD calibration in \cite{Osorio19b}, which is  currently considered the most computationally efficient approach.

Compared with \cite{Osorio19b}, we propose a novel method with enhanced computational efficiency and enhanced scalability (i.e.,  suitable for large-scale  road networks). This is achieved by using a simplified analytical network model defined as a system of linear, rather than nonlinear, equations and the dimension of which scales linearly with the number of roads with field measurements, as opposed to the total number of roads in the network. In the Salt Lake City case study of this paper, this represents a dimensionality reduction of two orders of magnitude: from tens of hundreds to hundreds of equations.  
Section~\ref{sec:method} presents the proposed methodology. Considering a large-scale Salt Lake City case study, we  benchmark the proposed approach versus two commonly used OD calibration methods showing the ability of the proposed method to quickly identify points with good performance (Section~\ref{sec:experiments}). 
Conclusions and future work are discussed in Section~\ref{sec:cl}.

\section{Methodology}
\label{sec:method}

\subsection{Problem formulation}
We consider the most classical OD calibration problem, which aims to identify an expected travel demand such as to replicate field measurements of vehicular count data on a set of roads. These measurements are the most commonly available type of traffic data collected by cities around the world.  The OD calibration is formulated as follows.
\begin{equation}
\min_{x \in \Omega} { f(x)  = \frac{1}{|I|} \sum_{i \in I} {(y_i - E[F_i(x,u_1;u_2])^2} + \delta \frac{1}{|Z|} 
\sum_{z \in Z} (x_z - \tilde{x}_z)^2}.
\label{eq:pbFormulation}
\end{equation}

The decision vector is represented by $x$ . The index set $Z$  is the set of integers in $[1,|Z|]$, where $|Z|$ represents the cardinality of $Z$. Element $x_z$ denotes the expected travel demand for the $z^\text{th}$ OD pair. The objective function $f$ represents  the sum of  2 main terms. The first summation represents the distance between the field measurements and their corresponding simulated counterparts. For a given road $i$  the field-measured vehicular count is denoted $y_i$, and the simulation-based expected count is denoted $E[F_i(x,u_1;u_2]$. It is denoted as the expectation of a random variable $F_i$ which represents the vehicular count on road $i$ and it depends on the decision vector $x$  as well as on a vector of endogenous simulation variables $u_1$ (e.g., road speeds, road travel times) and a vector of exogenous simulation parameters $u_2$ (e.g., road network topology, road attributes such as speed limits, number of lanes, curvature). The set of roads with field measurements is denoted $I$. Vehicular count data is typically obtained from underground loop detectors installed by cities on a sparse set of roads. Hence, the cardinality of $I$, denoted |I|, is typically orders of magnitude lower than the total number of roads. 
The second summation term is a regularization term that represents the squared distance between the decision vector and a prior vector, denoted $\tilde{x}$. The latter is known as a prior OD matrix or a seed matrix. It is often obtained from census data or travel surveys. The  parameter $\delta$  represents the weight assigned to the regularization term. The regularization term is needed because the problem is underdetermined. In other words, for a given set of roads with  count measurements available, there are typically an infinite number of vectors $x$ that can reproduce these counts. This is due both to the type of measurements (road counts, rather than more informative trajectory data) and to their sparsity (i.e., few roads have measurements). 
The feasible region $\Omega$ is available in analytical form and typically consists of a set of convex constraints. In this paper, we consider bound constraints. 

In this problem the objective function is not analytical, instead it is a function of simulation outputs (terms $E[F_i(x,u_1;u_2]$). Since the simulator is stochastic, the objective function can only be estimated by running various simulation replications, each of which is computationally costly to evaluate. The high computational cost of evaluating the simulator also limits the use of traditional deterministic gradient-based optimization techniques.  

\subsection{Main framework}
To tackle Problem~\eqref{eq:pbFormulation} this paper builds upon the general metamodel framework in \cite{Osorio13} and its specific formulation for OD calibration problems in \cite{Osorio19b}. 
The general idea of metamodel methods is to tackle Problem~\eqref{eq:pbFormulation} by solving a sequence of analytical problems that are obtained by replacing the simulation-based objective function with an analytical parametric function, known as the metamodel.  
More specifically, at a given iteration $k$ of the SO algorithm, all available simulation observations are used to fit the parameter vector $\beta_k$ of the metamodel and the following analytical problem is solved:
\begin{math} \min_{x \in \Omega} m_{k}(x;\beta_{k}) \end{math}. The solution of this problem is then simulated. As new simulation observations become available, the parameter of the metamodel is updated and new metamodel optimization problems are solved. Metamodel ideas were developed by the stochastic simulation community \cite{Ankenman10, Barton06}.
The most common choices are general-purpose functions, also known as functional metamodels, such as polynomials, splines, Kriging or Gaussian process  functions.  The main idea of \cite{Osorio13} was to consider specific classes of transportation problems and to formulate metamodels that embedded problem-specific structural information. In particular, novel analytical, differentiable, traffic network models were formulated  and used as part of the metamodel. The proposed metamodel methods preserve the generality and the asymptotic performance guarantees of the underlying SO algorithms,  while enhancing their computational efficiency for specific classes of transportation problems. 

For the OD calibration problem of \cite{Osorio19b}, the metamodel optimization problem is formulated, at a given iteration $k$,  as follows. 
\begin{equation}
\min_{x \in \Omega} { m_{k}(x;\beta_{k})  = \left( \beta_{k0} f^A(x) + \phi(x; \beta_{k}) \right) + \delta \frac{1}{Z} \sum_{z \in Z} ( x_{z} - \tilde{x}_{z} )^2 }
\label{eq:metam1}
\end{equation}
\begin{equation}
h(x,v;p)=0.
\label{eq:metam2}
\end{equation}
Problem~\eqref{eq:metam1}-\eqref{eq:metam2} differs from Problem~\eqref{eq:pbFormulation} in two ways. Firstly, the first summation term of~\eqref{eq:pbFormulation} is replaced by an analytical term. This analytical approximation is defined as the sum of a problem-specific approximation, represented by the function  $f^A$, and a general-purpose approximation, represented by the function  $\phi$. The latter is a parametric linear function of $x$, with parameter vector denoted $\beta_k$. The function $f^A$ is the analytical approximation of the first summation term of~\eqref{eq:pbFormulation}  as derived from an analytical traffic network model, which is defined by a system of equations and is denoted by $h$ in Equation~\eqref{eq:metam2}. 

The computational efficiency and the scalability of the metamodel framework, and of its extensions to specific classes of transportation problems, relies mainly on the formulation of a suitable function $h$. For OD calibration \cite{Osorio19a, Osorio19b}, the function $h$ is defined as a system of $n$ nonlinear equations, where $n$ is the number of roads in the network. Scalability is achieved because the complexity of the network model $h$ scales linearly with the number of roads in the network, and does not depend on the route choice set or on any road attributes (e.g., road length). 
The traffic models used account for endogenous route choice, meaning that the probability of choosing a specific route depends on the travel time of that route and of other routes in the network.  More generally, this allows the model to account for how congestion levels impact route choice. This is an important and realistic assumption. 
In this paper, we simplify the above-mentioned network model and propose a formulation for $h$ that consists of a system of equations that is: (i) of lower dimension, and (ii) linear, rather than nonlinear. This simplification enhances the computational efficiency of the overall approach. 

\subsection{Model formulation}
\label{sec:model}

Problem~\eqref{eq:metam1}-\eqref{eq:metam2} is formulated as follows.\\

\begin{multline}\label{eq:metam3}
\min_{x \in \Omega} { m_{k}(x;\beta_{k})}  = \beta_{k0} \frac{1}{|I|}  \sum_{i \in I} {(y_i - \lambda_i)^2}  + \beta_{k1} +
\sum_{z \in Z}  \beta_{k(z+2)} x_z  + \delta \frac{1}{Z} \sum_{z \in Z} ( x_{z} - \tilde{x}_{z} )^2     
\end{multline}

\begin{equation}
\lambda_i = \sum_{r \in R_i} { P_r x_{O(r)}} \ \ \forall i \in I.
\label{eq:metam4}
\end{equation}
The general-purpose function $\phi$ of Equation~\eqref{eq:metam1} is defined in~\eqref{eq:metam3} as a linear function of $x$ with  parameters ($\beta_{k1}, \beta_{k2}, ...\beta_{k(|Z|+2)}$).  The problem-specific approximation $f^A$ of ~\eqref{eq:metam1} is defined as a sum of squares that measures the distance between the field measured vehicular count $y_i$ and the expected vehicle count derived from the analytical network model, denoted $\lambda_i$ and defined by~\eqref{eq:metam4}. 
In  Equation~\eqref{eq:metam4}, $R_i$ denotes the set of routes that travel through road $i$, $O(r)$ denotes the OD pair of route $r$ and $P_r$ denotes the probability of choosing route $r$. The right-hand side of  Equation~\eqref{eq:metam4} represents the total expected demand for road $i$, it is defined as the sum of the expected  demand for all routes that include road $i$.
Equation~\eqref{eq:metam4} approximates the expected flow on road $i$, $\lambda_i$, with the expected demand for road $i$. 
The system of linear equations~\eqref{eq:metam4} represents the function $h$ of \eqref{eq:metam2}. 
The work of  \cite{Osorio19b} used instead a system of nonlinear equations. The nonlinearity stems from the fact that the route choice probability was considered endogenous, i.e., $P_r$ was a function of $x$. More specifically, the choice probability of a given route is assumed to be a function of the route's travel time. This is a realistic assumption. In our work, we use the Google Maps Distance Matrix API to obtain an accurate travel time estimate.
We then compute the choice probability of a route $r$, $P_r$, based on a multinomial logit choice model just as in \cite{Osorio19b}: 
\begin{math} P_r = \frac{\exp^{\theta t_r}}{ \sum_{j \in R_2(r)}{\exp^{\theta t_j}} } \end{math}, where  $t_j$ denotes the travel time of a given  route $j$, $R_2(r)$ denotes the set of routes that share the same OD pair as route $r$, and $\theta$ is a travel time scalar parameter.  The use of exogenous time-of-day-specific travel times allows us to estimate $P_r$ exogenously. This greatly simplifies the system of equations of  \cite{Osorio19b} and leads to the linear system of equations defined in~\eqref{eq:metam4}.  

\begin{figure}[h]
  \centering
\includegraphics[width=\linewidth]{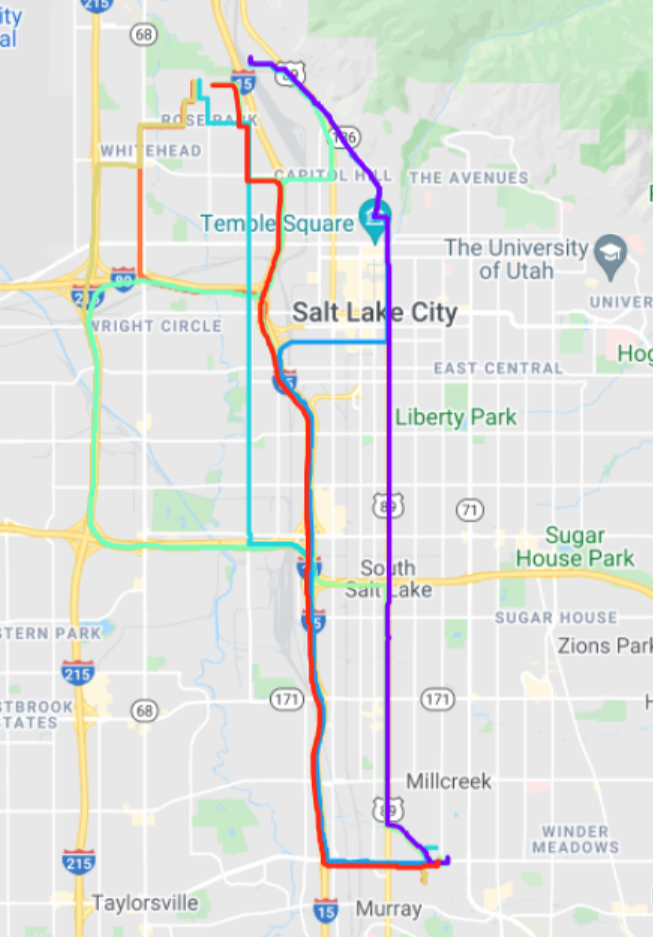}
  \caption{Sample set of alternate routes queried from Google Maps API for a particular OD. The routes cover diverse roads and are competitive in terms of travel time.}
  \label{fig:slcNwk}
  \end{figure}

The proposed formulation has enhanced computational efficiency compared to  \cite{Osorio19b}  for two reasons: (i) it consists of a system of linear, rather than nonlinear equations, (ii) the dimension of the  system of equations is the number of roads with field measurements, $|I|$, rather than the total number of roads in the network, $n$. Typically,  $|I|$ is one to two orders of magnitude lower than $n$. In the case study of this paper, this yields a reduction  of two orders of magnitude (45410 versus 429). 
Moreover, we implement  Problem~\eqref{eq:metam3}-\eqref{eq:metam4} as a bound constrained analytical problem with a quadratic objective function, for which there are a variety of efficient solvers.

Equation~\eqref{eq:metam4} can be rewritten in matrix format as
\begin{math} \lambda = \tilde{P} x \end{math}. In this case, the matrix $\tilde{P}$ is known in the literature as the linear assignment matrix. It maps the OD demand to the road counts. A review of OD  calibration methods that rely on the estimation and use of a linear assignment matrix is given in \cite{Toledo13}. Most approaches iteratively estimate $\tilde{P}$ via simulation. However, this comes at increased  computational cost, since the estimation of $\tilde{P}$ requires solving a simulation-based fixed-point problem. As discussed in \cite{Toledo13}, a variety of approaches have been proposed to reduce the computational cost of this estimation. Such a method is used as a benchmark in the numerical experiments of Section~\ref{sec:experiments}. Our proposed approach differs from linear assignment methods in two ways. First, it provides an accurate and exogenous estimation of  $\tilde{P}$, as opposed to resorting to simulation-based estimation. This is achieved through the use of accurate time-of-day specific route travel time estimations obtained from the Google Maps Distance Matrix API. 
Second, our objective function is a  metamodel approach that parametrically corrects for the analytical approximation provided by the analytical network model, while traditional linear assignment methods do not do so (i.e., they set $\beta_{k0}=1$ and $\beta_{ki}=0 \  \forall i \geq 1$).

\section{Salt Lake City Experimental Results}
\label{sec:experiments}

\subsection{Experimental design}

The network of interest is that of Salt Lake City. Figure~\ref{fig:slcNwk} displays the corresponding road network.
Summary statistics of the size of the network and problem are given in Table~\ref{tab:nwk}.
The time period of interest is 5-6pm on December 9, 2020. 
We use detailed maps data from Google Maps to define the road network topology and road attributes (e.g.,  length, curvature, lanes, speed-limits, priority level, control type). 
 The set of OD pairs as well as the prior (or seed) OD is defined by using the synthetic population generation method described in \cite{Abueg21}, which builds off of the work of \cite{Beckman96,Barrett09}, and combines  2010 US Census of Population and Housing data \cite{census10} and 2012-2016 ACS Public Use Microdata Sample (PUMS) data \cite{PUMS20}. 

The route set is constructed by querying Google Maps Directions API to obtain 10 plausible routes per OD pair, along with their estimated travel time for the specific time of interest. For each OD pair, we ensure route choice diversity  by limiting the distance-based overlap of any pair of routes to 70\%.  The model uses the traffic simulation software SUMO \cite{SUMO_2018}. 
The field measurements consist of  hourly counts for the the specific time period obtained from the 
Utah Department of Transportation (UDOT) Automated Traffic Signal Performance Metrics (ATSPM) portal \cite{udotAtspm}. We use measurements on 429 roads (i.e., $|I|$=429). 

\begin{table}
\begin{center}
    \begin{tabular}{|c|c|c|}
    \hline
    \# OD pairs & \# edges & \# routes \\
    \hline
    31937 & 45410 & 260837   \\
    \hline
    \end{tabular}
\end{center}
\caption{Summary statistics of the road network topology}
\label{tab:nwk}
\end{table}

\begin{figure}[h]
  \centering
\includegraphics[width=\linewidth]{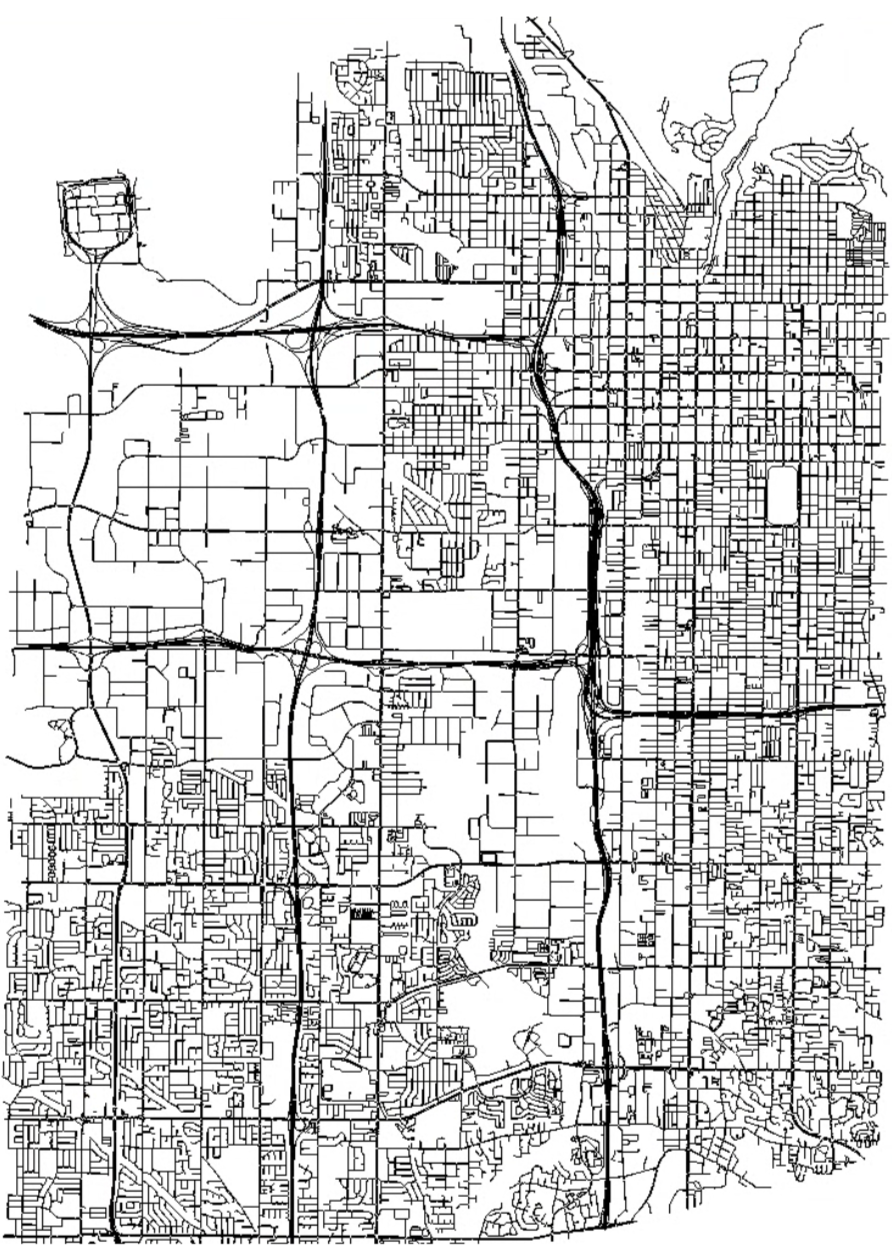}
  \caption{Topology of the Salt Lake City road network model}
  \label{fig:slcNwk}
  \end{figure}

We benchmark the proposed approach against the following methods.
\begin{itemize}
    \item Simultaneous Perturbation Stochastic Approximation (SPSA) \cite{Spall92,Spall03}. As discussed in Section~\ref{sec:intro}, this is a commonly  used algorithm  for OD calibration. 
    The hyperparameters are set based on various guidelines (\cite{Balakrishna06,Vaze09} and Chapter 7 of \cite{Spall03}) as well as values used in the Singapore case study of \cite{Lu15}. The values used, based on ODs in veh/hr units and using the parameter notation of \cite{Spall03}, are:  $\alpha=0.602, \gamma = 0.101, c=1.9, a=0.16, A= 0.02$.

    \item Linear Assignment Method (LAM). We use the method of \cite{Toledo04} that is based on the commonly used linear assignment methods discussed in Section~\ref{sec:model}. 
    We use a gradient descent based solver implemented using Tensorflow~\cite{Abadi15} for solving the quadratic minimization problem. The learning rate used is 0.001. The estimated assignment matrix is updated after each  simulation iteration (or simulation round) using the method of successive averages as follows.
    \begin{equation}
        A^{t+1} = (1 - \frac{1}{t})  A^{t} + \frac{1}{t}  \hat{A}^{t+1},
    \end{equation}
    where $t$ is the simulation iteration;  $A^{t}$ is the estimated assignment matrix  of  iteration $t$; and $\hat{A}^{t+1}$ is the simulation-based estimate of the assignment matrix obtained from the simulation run of iteration $t+1$. 
    
\end{itemize}

We refer hereafter to the proposed method as the Linear-Metamodel method.  For all methods, we use the prior OD matrix, estimated from the generated synthetic population mentioned above, as the initial point. All methods are terminated after $15$ SO iterations. 

\subsection{Numerical results}

 We use the  normalized root mean square error (nRMSE) as a summary statistic:  
\begin{equation}
    nRMSE = 100 * \frac{\sqrt{\frac{1}{|I|} \sum_{i \in I} {(y_i - \hat{E}[F_i(x,u_1;u_2])^2}}}{\frac{1}{|I|} \sum_{i \in I} {y_i}},
\end{equation}
where $\hat{E}[F_i(x,u_1;u_2]$ denotes the simulation-based estimate of $E[F_i(x,u_1;u_2]$.

\begin{figure}[h]
    \centering
    \includegraphics[width=\linewidth]{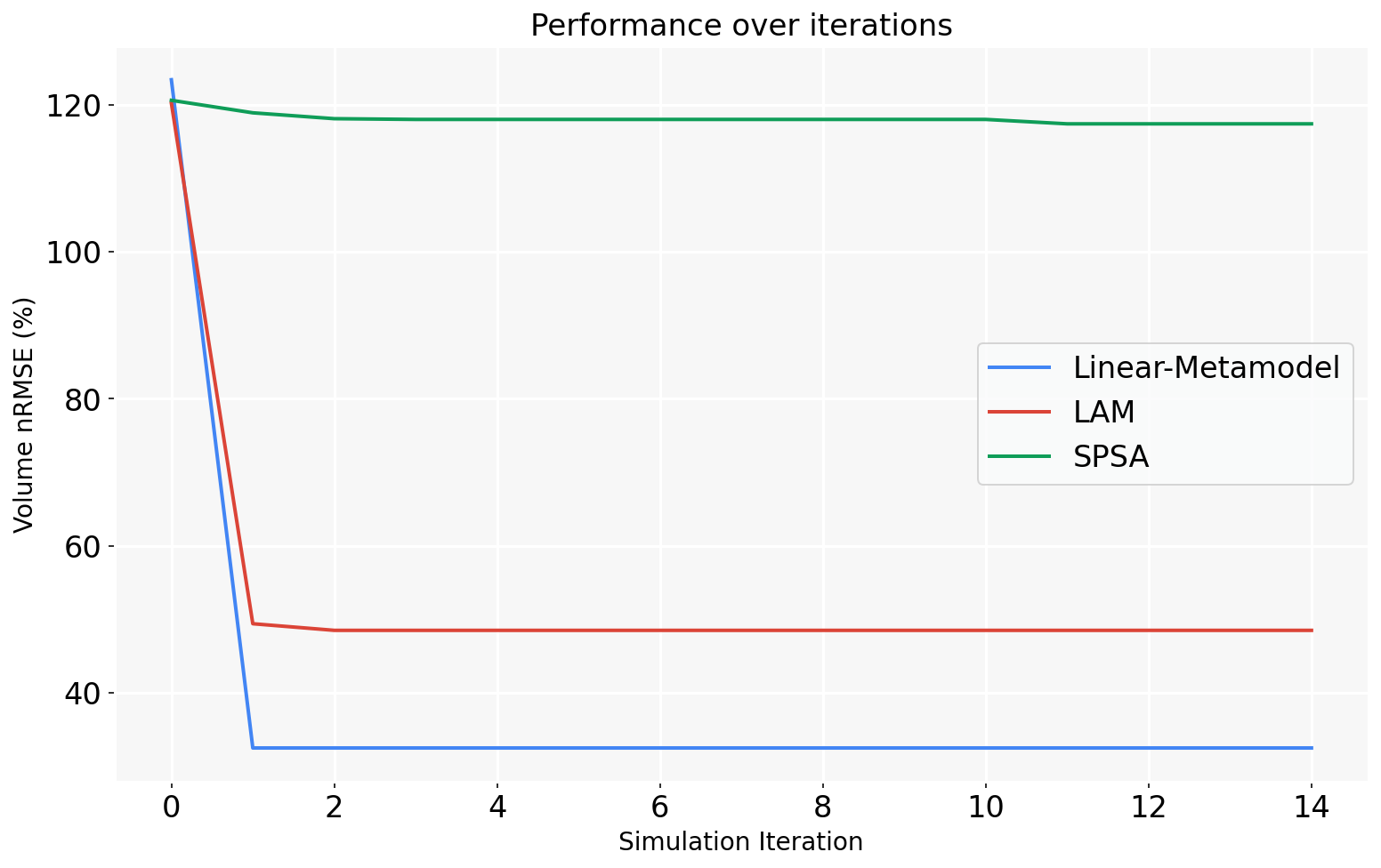}
    \caption{Comparison of the objective function of the current iterate of SPSA, LAM and (proposed) Linear-Metamodel  as a function of  simulation iterations.}
    \label{fig:iterationsComp}
\end{figure}

Figure~\ref{fig:iterationsComp} displays the nRMSE of the current iterate (point with  best simulated performance). SPSA gradually identifies points with small improvements in  performance.  This is  expected because  it is not designed to be terminated within few simulation iterations. OLS and Linear-Metamodel identify, at the first iteration, ODs with substantially improved performance. Thereafter there is not much improvement within such few iterations. The nRMSE of the intial OD is 120\%, that of the best point identified by the  Linear-Metamodel is  32\%,  that of LAM is 48\% and that of SPSA is 117\%. LAM and the Linear-Metamodel outperform SPSA. The Linear-Metamodel  reduces the nRMSE by 33\% compared to  LAM. 

\begin{figure*}[ht]
\begin{center}
  \centering
\scalebox{0.5}{
\includegraphics[width=\textwidth]{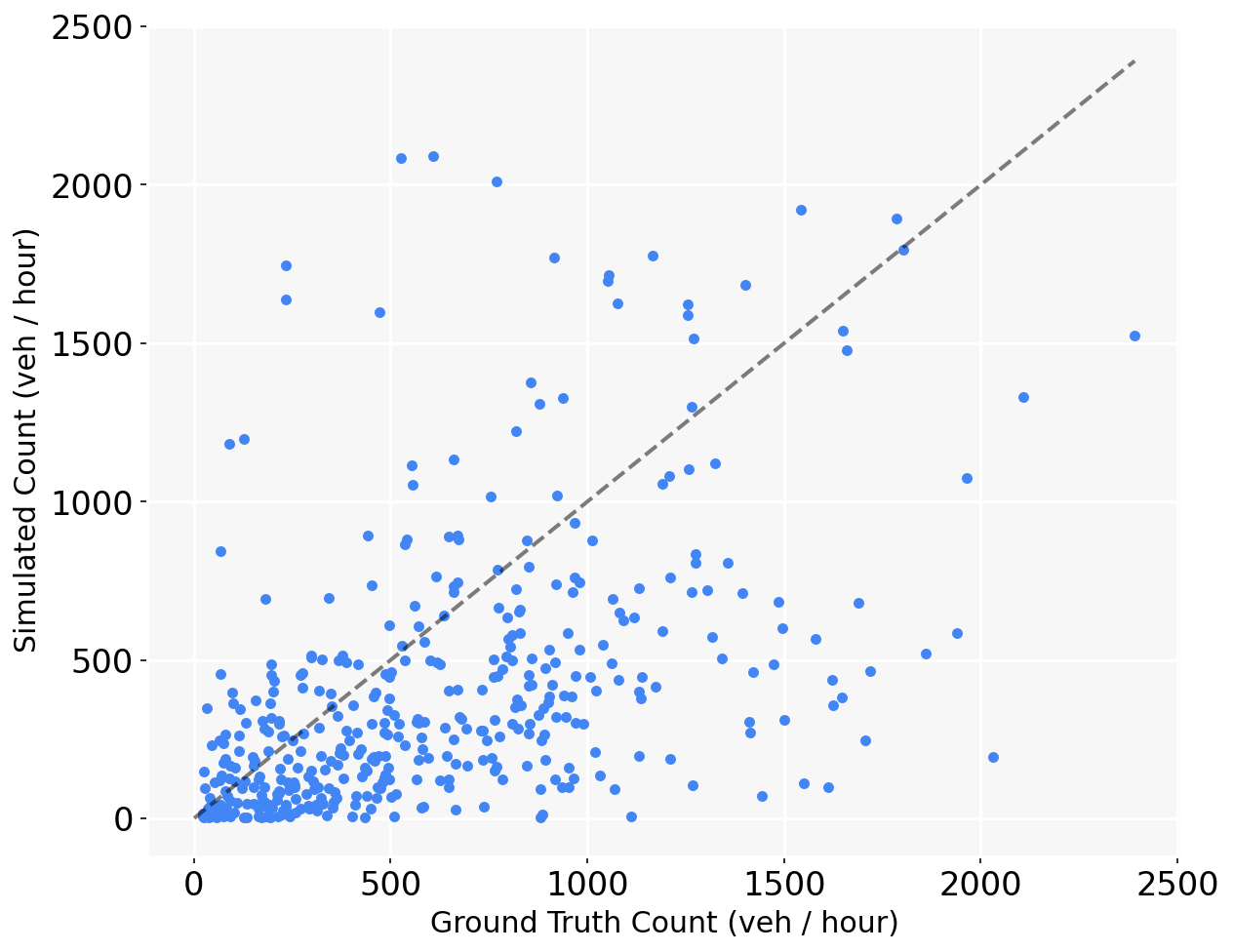}
\includegraphics[width=\textwidth]{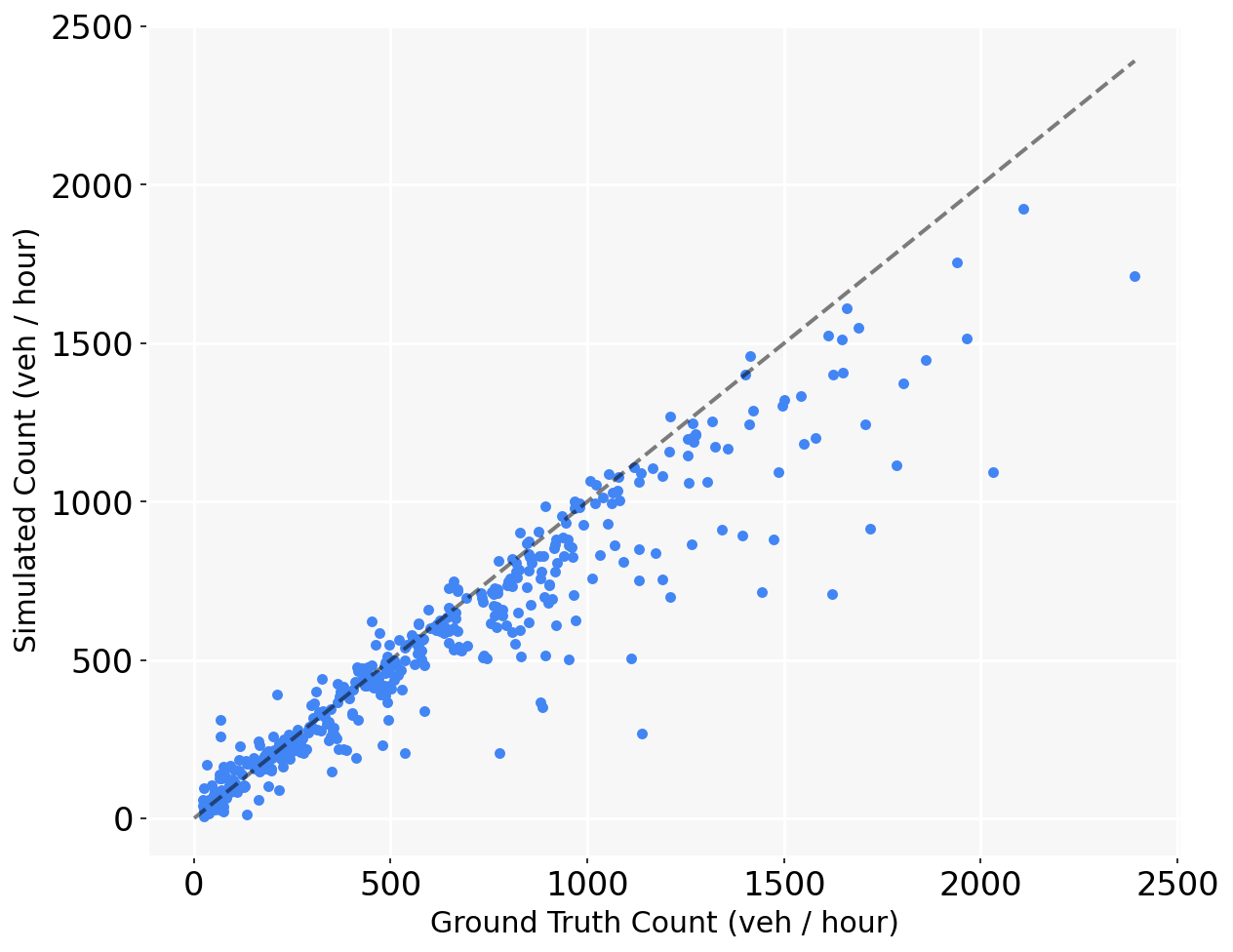}
}\\
\scalebox{0.5}{
\includegraphics[width=\textwidth]{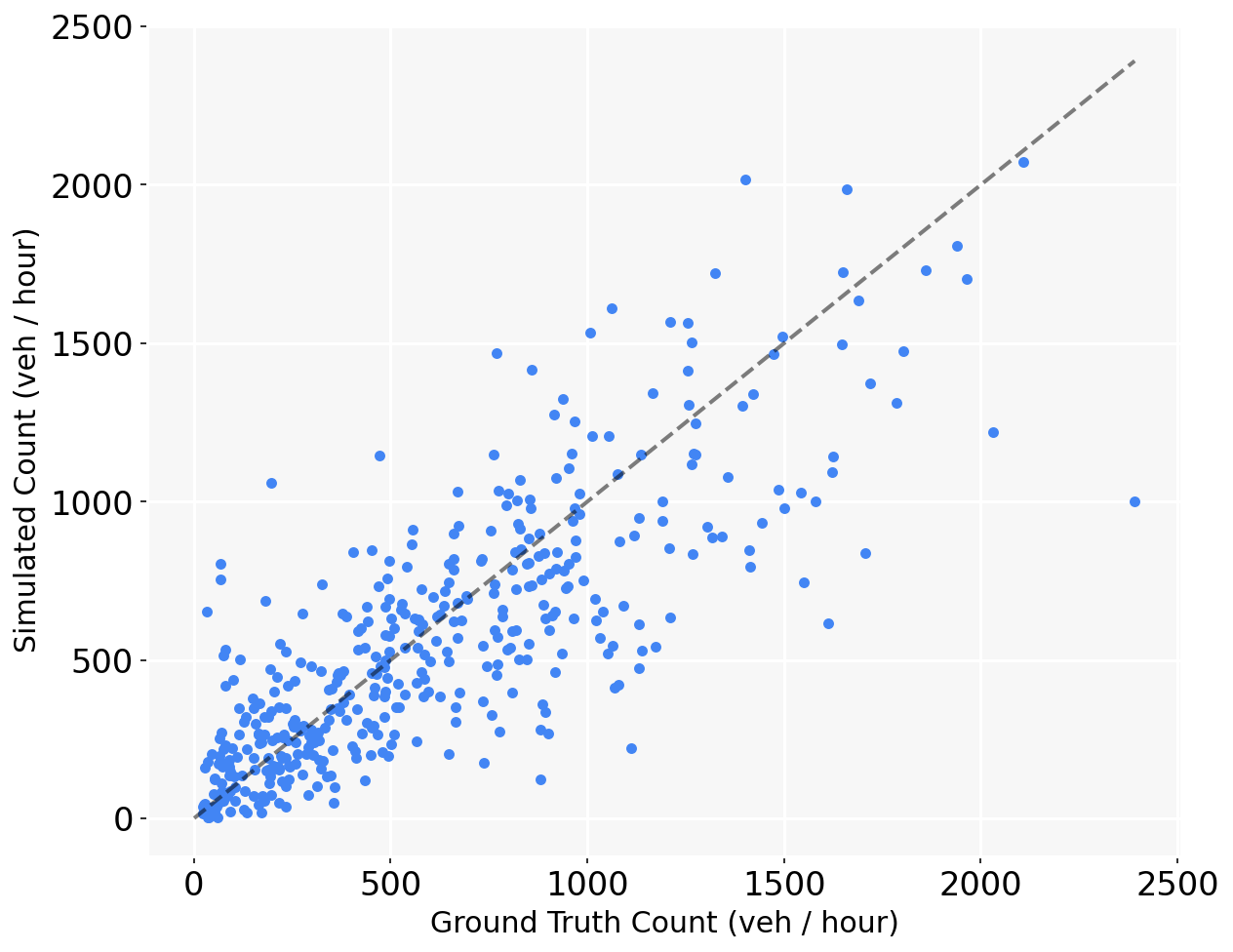}
\includegraphics[width=\textwidth]{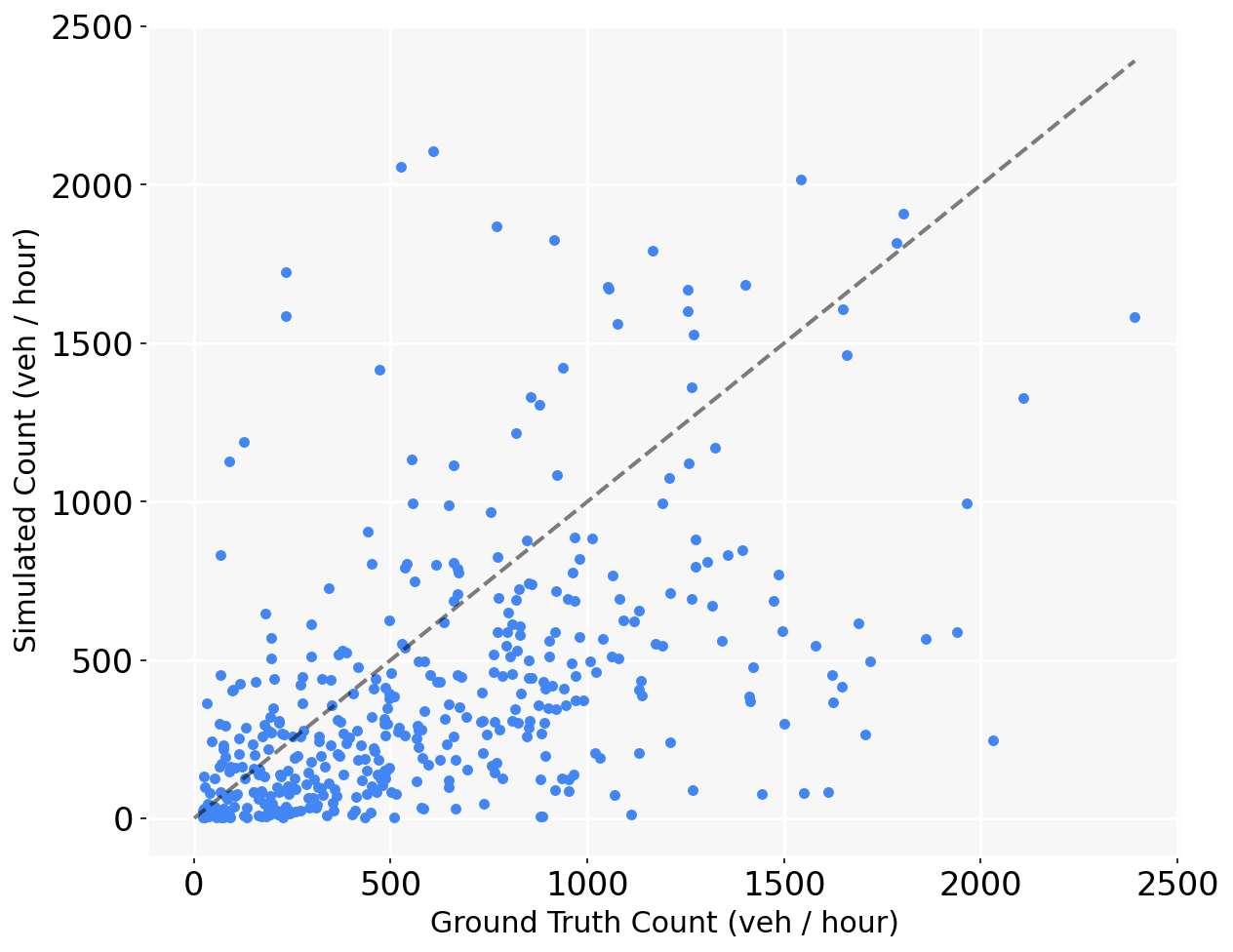}
}
  \caption{Comparison of field counts and simulated counts. The top left plot considers the initial OD, the top right corresponds to the best OD identified by Linear-Metamodel, the bottom left and bottom right are best ODs identified by LAM and SPSA respectively.}
  \label{fig:scatterPropLAM}
\end{center}
\end{figure*}

Figure~\ref{fig:scatterPropLAM} displays the fit to the field data for the  best OD found by each method.  For each plot, the x-axis displays the  field measurements,  i.e., $y_i$ of Equation~\eqref{eq:pbFormulation}. The y-axis displays the corresponding simulated estimates, i.e., the estimates of $E[F_i(x,u_1;u_2]$ of Equation~\eqref{eq:pbFormulation}. The more the points fall along the diagonal dashed line, the better the fit to the field data. These plots show a more detailed analysis of the fit to the field data, compared to the previous more aggregate  nRMSE analysis. These plots illustrate the similarity of the fit to the field data between the OD of SPSA and the initial OD (i.e., top left plot and bottom right plot). The improvement in fit to field data between Linear-Metamodel (top right plot) and LAM (bottom left plot) is also clear.

\section{Conclusions}
\label{sec:cl}

This paper proposes a metamodel technique to tackle a class of high-dimensional continuous simulation-based optimization problems known as OD calibration problems. The proposed method builds off of an existing metamodel approach, which resorts to  solving a sequence of approximate analytical optimization problems with nonlinear constraints, the dimension of which scales linearly with the number of roads in the network. The use of detailed exogenous travel time data allows us to propose a formulation that solves a problems with linear constraints, the dimension of which scales linearly with the number of roads with field data. This translates, in the Salt Lake City case study of this paper, to a reduction in the number of constraints of two orders of magnitude. The method outperforms a traditional general-purpose SO algorithm, SPSA, as well as a commonly used and tailored algorithm for OD calibration (linear assignment method).  Future work includes the extension of this method to tackle dynamic OD calibration problems, and to allow for other types of field data (e.g., road speeds, trajectory  data).

\bibliography{refs}
\end{document}